\documentclass[twocolumn,amsmath,amssymb, 14pt]{revtex4}
\usepackage{amsmath}
\usepackage{amssymb}
\usepackage{amsmath,amsthm,amssymb,amscd}
\usepackage{latexsym}
\usepackage{indentfirst}
\usepackage{subfigure}
\usepackage{graphicx}

\begin{document}

\title{
  \bf Comment on \lq\lq Speed Limit for Classical Stochastic Processes"
}

\author{Yunxin Zhang}{
 \affiliation{ School of Mathematical Sciences and Centre for Computational Systems Biology, Fudan University, Shanghai 200433, China.}

\begin{abstract}
In recent letter [Phys. Rev. Lett {\bf 121}, 070601 (2018)], the speed limit for classical stochastic Markov processes is considered, and a trade-off inequality between the speed of the state transformation and the entropy production is given. In this comment,  a more accurate inequality will be presented.
\end{abstract}

\maketitle

In recent letter of Shiraishi {\it et al.} \cite{Shiraishi2018}, for classical stochastic Markov processes, a trade-off inequality between the speed of the state transformation and the entropy production is presented. This result is very important for the study of optimization of heat engines, or even general stochastic Markov processes \cite{Gingrich2016,Pietzonka2018,Solon2018}. In this comment, by using a new mathematical inequality, I will give an improved result.

For a stochastic Markov process with discrete states governed by the following equation,{\small
\begin{equation}\label{eq1}
\begin{aligned}
\frac{d}{dt}p_i(t)=\sum_{j(\ne i)}[W_{ij}(t)p_j(t)-W_{ji}(t)p_i(t)],
\end{aligned}
\end{equation}}
the total variation distance between $\textbf{p}(0)$ and $\textbf{p}(\tau)$ is defined as{\small
\begin{equation}\label{eq2}
\begin{aligned}
L(\textbf{p}(0), \textbf{p}(\tau))=\sum_{i}|p_i(0)-p_i(\tau)|.
\end{aligned}
\end{equation}}
Where the transition rate matrix satisfies the normalization condition {\small $\sum_{i}W_{ij}(t)=0$} and non-negativity {\small $W_{ij}(t)\ge 0$} for $i\ne j$.
In \cite{Shiraishi2018}, it is shown that{\small
\begin{equation}\label{eq3}
\begin{aligned}
\tau_I:=\frac{L(\textbf{p}(0), \textbf{p}(\tau))^2}{2\Sigma\langle A\rangle_\tau}\le\tau.
\end{aligned}
\end{equation}}
Where {\small $\Sigma:=\int_0^\tau dt\dot{\Sigma}(t)$}, and {\small $\langle A\rangle_\tau:=\frac1\tau\int_0^\tau dt A(t)$} with{\small
\begin{equation}\label{eq4}
\begin{aligned}
&\dot{\Sigma}(t)=\frac12\sum_{i}\sum_{j(\ne i)}\left[(W_{ji}p_i-W_{ij}p_j)\ln\frac{W_{ji}p_i}{W_{ij}p_j}\right],\cr
&A(t)=\sum_{i}\sum_{j(\ne i)}[W_{ji}(t)p_i(t)].
\end{aligned}
\end{equation}}

I found that, instead of the lower bound $\tau_I$ of $\tau$ as given in (\ref{eq3}), there exists a more accurate one,{\small
\begin{equation}\label{eq5}
\begin{aligned}
\hat{\tau}_I:=\frac{2L(\textbf{p}(0), \textbf{p}(\tau))^2}{\Sigma\langle B\rangle_\tau}\le\tau,
\end{aligned}
\end{equation}}
in which {\small $\langle B\rangle_\tau:=\frac1\tau\int_0^\tau dt B(t)$} with
{\small $B(t):=\sum_{i}B_i(t)$} and {\small  $B_i(t):=\sum_{j(\ne i)}\left[\sqrt{W_{ji}(t)p_i(t)}+\sqrt{W_{ij}(t)p_j(t)}\right]^2$}.
Obviously {\small $\langle B\rangle_\tau\le 4\langle A\rangle_\tau$}, therefore $\tau_I\le \hat{\tau}_I\le \tau$. The derivation of inequality  (\ref{eq5}) is as follows.
{\small
\begin{eqnarray}\label{eq6}
&&L(\textbf{p}(0), \textbf{p}(\tau))\le\sum_{i}\int_0^{\tau}dt\left|\frac{d}{dt}p_i\right|\cr
&=&\sum_{i}\int_0^{\tau}dt\left|\sum_{j(\ne i)}[W_{ij}(t)p_j(t)-W_{ji}(t)p_i(t)]\right|\cr
&\le&\int_0^{\tau}dt\sum_{i}\sqrt{B_i(t)S_i(t)}
\le\int_0^{\tau}dt\sqrt{B(t)S(t)}\cr
&\le&\left(\int_0^{\tau}dtB(t)\right)^{1/2}
\left(\int_0^{\tau}dtS(t)\right)^{1/2}\cr
&\le&\sqrt{\tau \langle B\rangle_{\tau}\Sigma/2},
\end{eqnarray}}
where {\small $S_i(t):=\sum_{j(\ne i)}\left[\sqrt{W_{ij}(t)p_j(t)}-\sqrt{W_{ji}(t)p_i(t)}\right]^2$} and {\small $S(t):=\sum_{i}S_i(t)$}. In the third and forth lines of (\ref{eq6}), the Schwarz inequality is used, and in the last line the inequality {\small $(a-b)\ln(a/b)\ge4(\sqrt{a}-\sqrt{b})^2$} is used. From (\ref{eq6}), inequality (\ref{eq5}) can be obtained easily.

For the example discussed in \cite{Shiraishi2018}, transition rates {\small $W_{10}=1$, $W_{01}=(4\tau+1)/(2\tau-t)-1$}, probability {\small $p_1(t)=1/2-t/(4\tau)$, $p_0(t)=1/2+t/(4\tau)$}. Which give that {\small $\langle A\rangle_\tau=5/4+1/(4\tau)$, $L(\textbf{p}(0), \textbf{p}(\tau))=1/2$, $\Sigma=[(3\tau+1)\ln(3\tau+1)-(2\tau+1)\ln(2\tau+1)-3\tau\ln(3\tau)+2\tau\ln(2\tau)]/(4\tau)$}, and {\small $\langle B\rangle_\tau=2\langle A\rangle_\tau+4C$} with {\small $C=1/(4\tau^2)\int_0^\tau dt\sqrt{(t+2\tau+1/2)^2-1/4}$}. Obviously, {\small $4C<1/\tau^2\int_0^\tau dt(t+2\tau+1/2)=5/2+1/(2\tau)=2\langle A\rangle_\tau$}, therefore {\small $\langle B\rangle_\tau<4\langle A\rangle_\tau$} and consequently $\tau_I<\hat{\tau}_I$. See Fig. \ref{fig} for the plots of difference $\hat{\tau}_I-\tau_I$, as well as the relative errors $(\tau-\hat{\tau}_I)/\tau$ and $(\tau-\tau_I)/\tau$.

\begin{widetext}

\begin{figure}
  \includegraphics[width=8.5cm]{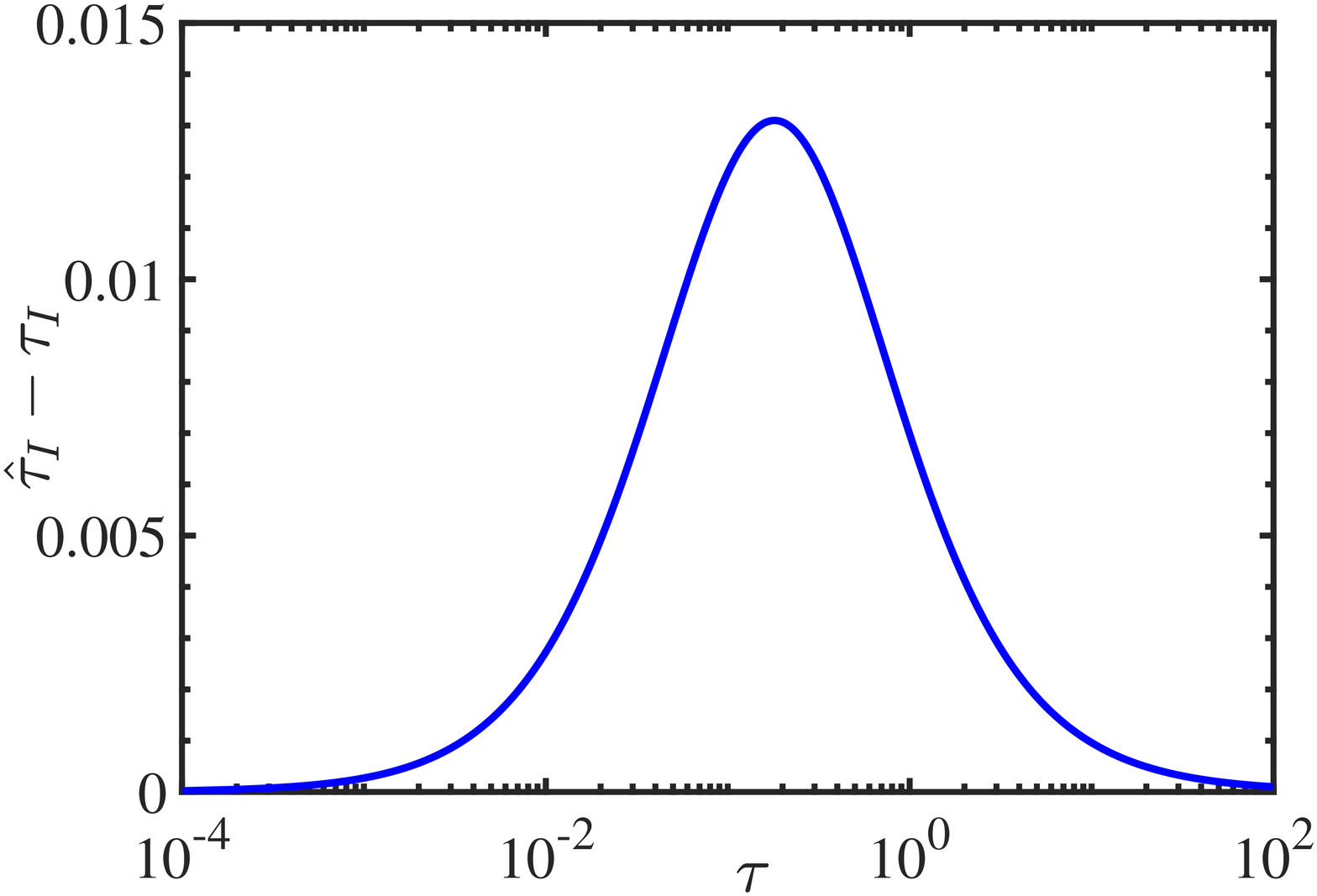}\includegraphics[width=8.5cm]{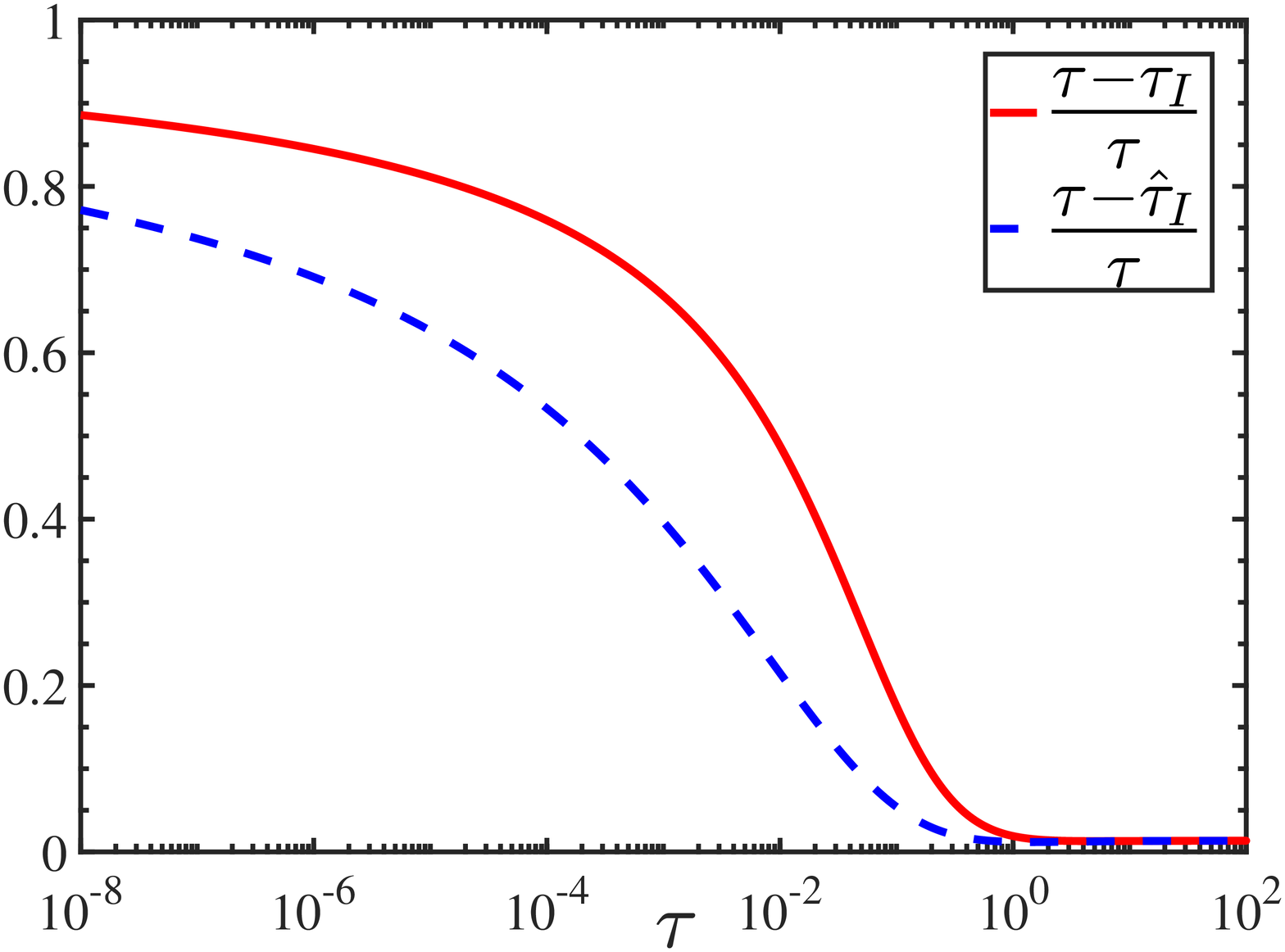}\\
  \caption{The difference $\hat{\tau}_I-\tau_I$ between the improved lower bound $\hat{\tau}_I$ obtained in this comment and the lower bound $\tau_I$ obtained in \cite{Shiraishi2018} \textbf{(left)}, and the relative errors $(\tau-\hat{\tau}_I)/\tau$ and $(\tau-\tau_I)/\tau$ \textbf{(right)}. }\label{fig}
\end{figure}

\end{widetext}

\end{document}